\documentclass[10pt, twocolumn]{IEEEtran}
\normalsize
\usepackage{amsthm,amssymb,amsmath, tikz,graphics,cite,float,graphicx,epstopdf,epsfig,verbatim,url}
\usetikzlibrary{automata}
\usetikzlibrary{shapes,arrows}

\usepackage[utf8]{inputenc}

\usepackage{capt-of}
\usepackage[noend]{algpseudocode}
\usepackage{algorithmicx}
\usepackage[ruled]{algorithm}

\usepackage{lipsum}

\usepackage{mdwtab}
\usepackage{subfigure}
\usepackage{placeins}
\usepackage{setspace}

\allowdisplaybreaks

\theoremstyle{plain}
\theoremstyle{remark}

\begin{document}
\newtheorem{theorem}{Theorem}[section]
\newtheorem{lemma}{Lemma}
\newtheorem{conjecture}{Conjecture}
\newtheorem{corollary}{Corollary}
\newtheorem{definition}{Definition}
\newtheorem{property}{Property}
\newtheorem{remark}{Remark}

\bibliographystyle{plain}

\title{Leveraging Edge Caching in NOMA Systems with QoS Requirements}

\author{\IEEEauthorblockN{Jos\'{e} Armando Oviedo and  Hamid R. Sadjadpour}\\
\IEEEauthorblockA{Department of Electrical Engineering, University of California, Santa Cruz\\
Email: \{xmando, hamid\}@soe.ucsc.edu}}

\maketitle

\IEEEpeerreviewmaketitle

\begin{abstract}
	Non-Orthogonal Multiple Access (NOMA) and caching are two proposed approaches to increase the capacity of future 5G wireless systems. Typically in NOMA systems, signals at the receiver are decoded using successive interference cancellation in order to achieve capacity in multi-user systems. The leveraging of caching in the physical layer to further improve on the benefits of NOMA is investigated, which is termed cache-aided NOMA. Specific attention is given to the caching cases where the users with weaker channel conditions possess a cache of the information requested by a user with a stronger channel condition. The probability that any of the users is in outage for any of the rates required for this NOMA system, defined as the  "union-outage," is derived for the case of fixed-power allocation, and the power allocation strategy that minimizes the union-outage probability is derived. Simulation results confirm the analytical results, which demonstrate the benefits of cache-aided NOMA on reducing the union-outages probability.

\end{abstract}

\section{Introduction}
Data caching and Non-Orthogonal Multiple Access (NOMA) are being  heavily investigated for implementation in the next generation of wireless systems, namely 5G \cite{Li-2014, BBD:ProCa}. NOMA has relied on superposition coding (SC) at the transmitter, and successive interference cancellation (SIC) at the receiver \cite{InfTh:CT,FundWiCom:Tse} in order to achieve capacity in multi-user wireless systems. Given the advanced storage and processing capabilities of current and future user terminals, caching can then be leveraged at the receiver side of a user with weaker channel strength in a NOMA system.  

Consider a NOMA system with caching enabled user terminals. While a user with stronger channel can use SIC to decode and remove the interfernce caused by the signal intented for the user with weaker channel, the user with weaker channel can take advantage of caching if it has already cached the information requested by the user with stronger channel. Since the header of the signal contains the modulation and rate information for each signal to be decoded, and can contain a unique file identifier for the data, the user with weaker channel can then recreate the signal transmitted for the stronger user and subtract it from the composite signal, similar to how successive interference cancellation works. The difference here is that the rate at which the information of the other user is sent does not have to be within the capacity of the user containing the cache. Thus, this technique called cache-aided NOMA (CA-NOMA) creates two interference free channels for the two users by taking advantage of two proposed techniques in the emerging 5G wireless standards. 

However, since the capacity of the weaker user's channel increases, this creates different problems related to the limits of what the stronger user can successfully decode using SIC. With regular NOMA, the capacity at which the weaker user's information is sent is always upper-bounded by the capacity at which the stronger user can perform SIC. With caching, this is no longer the case. 

The rest of the paper is organized as follows: section \ref{sec:previous} covers the relevant previous work on NOMA and caching, section \ref{sec:system} details the system model used in our analysis, section \ref{sec:QoS} outlines the definition of the union-outage event that occurs in NOMA and covers the analysis for satisfying a minimum QoS, section \ref{sec:results} compares our analysis to simulation results, and we share our concluding thoughts and propose future work in section \ref{sec:conclusion}.

\section{Previous Work on NOMA} \label{sec:previous}

The basic concept of NOMA was  described in \cite{InfTh:CT}, which showed that the approach of using SC at the transmitter and SIC at the receiver with greater signal-to-interference-plus-noise ratio (SINR) achieves capacity of the channel. It is known \cite{FundWiCom:Tse} that for every orthogonal approach to scheduling multiple users in a point-to-multipoint transmission, there always exists a non-orthogonal approach that will outperform it.  The exact power allocations and associated expected capacity gains were obtained in \cite{FairNOMAInfocom2016} for the 2-user SISO channel with completely random user-pairing. Then the results were extended in \cite{FairNOMAFull}  to the general $K$-user SISO systems to find the expected capacity gains for optimal user-pairing (according to \cite{5GNOMA:DFP}), and closed-form expressions were found for the power allocation coefficients that allow all users to achieve capacities greater than or equal to their respective OMA capacities. Similarly, in \cite{DNOMA:YDFA} the authors use a dynamic power allocation strategy that always outperforms OMA in order to improve the outage capacity performance of the network.

Caching at the edge of networks, including small Base Stations (BSs) and device-to-device communications for future 5G wireless networks is described in \cite{GreenCache}, where it was shown that caching is an energy efficient solution to help the network improve coverage probabilities while maximizing spectral efficiency. In \cite{FemCache:Mohsen}, it is demonstrated that using decentralized coded caches at user terminals increases the throughput capacity of the network, while alleviating the burden of network traffic caching at BSs and helper nodes.

In CA-NOMA, caching is used at the user terminals (users) in order to assist in eliminating interference from channels that cannot otherwise take advantage of SIC due to weaker channel conditions. We show that downlink capacity and outage probabilites improve when a user possesses a cache of information requested by a user with a stronger channel, then derive the fundamental power allocation constraints, and find the optimum power allocation that minimizes the union-outage probability  of the NOMA system.

\section{CA-NOMA System Model and Capacity} \label{sec:system}
Consider a two-user orthogonal multiple access (OMA) single-input-single-output (SISO) system. Let the total transmit time period be $T$, where users are allocated equal non-overlapping time slots of length $T/2$, and allocated the total transmit SNR $\xi$ for that slot\footnote{The same OMA formulation can be made in the case of frequency division instead of time division.}. For each user $k=1,2$, if the BS is transmitting signal $x_k$ ($\mathbb{E}[|x_k|^2]=1$), the channel gain between the BS and user-$k$ is $h_k$ ($|h_k|^2\sim\mathrm{Exponential}(\frac{1}{\beta})$), and the receiver has noise $z_k\sim\mathcal{CN}(0,1)$, then the received signal at user $k$ is given by 
\begin{align}
	\label{eq:OMAsignal}y_k = \sqrt{\xi}h_k x_k + z_k. 
\end{align}
Since the time duration is $\frac{1}{2}$, then the capacity of user $k$ in bps/Hz is given by
\begin{align}
	\label{eq:OMAcapacity}C_k^\text{oma} = \frac{1}{2}\log_2(1+\xi|h_k|^2).
\end{align}

Let the user with weaker channel be user-1, so that $|h_1|^2<|h_2|^2$. In a two-user SISO NOMA system, where the BS is transmitting the two signals using SC, the signal $x_i$ is carrying the information for user-$i$. The power allocation coefficients for users 1 and 2 are $1-a$ and $a$, respectively. The SC signal transmitted by the BS is then $\sqrt{(1-a)\xi}x_1+\sqrt{a\xi}x_2$, thus the raw received signal at user-$i$ is
	\begin{align}
	\label{eq:NOMAsignalRAW}r_i = h_i(\sqrt{(1-a)\xi}x_1+\sqrt{a\xi}x_2) + z_i, i=1,2.
	\end{align}
Since NOMA allows user-2 to remove the interference received by user-1's signal, the received signal after using SIC at user-2  is
\begin{align}
	&y_2 = \sqrt{a\xi}h_2x_2 + z_2. 
\end{align}
However, in a CA-NOMA enabled system where user-1 caches  user-2's information, user-1's receiver can decode user-2's information and subtract its signal from the composite received signal, and thus has its information carried on the signal
\begin{align}
	&y_1 = \sqrt{(1-a)\xi}h_1x_1 + z_1.
\end{align}
This gives rise to the capacities
	\begin{align}
		\label{eq:NOMAcap1_cache1}&C_1(a) = \log_2(1+ (1-a)\xi|h_1|^2)\\
		\label{eq:NOMAcap2_cache1}&C_2(a) = \log_2(1+ a\xi|h_2|^2). 
	\end{align}
The capacity for user-$2$ to decode and subtract user-$1$'s signal using SIC is 
\begin{align}
	\label{eq:NOMAcap21SIC}C_{2\rightarrow 1}(a) = \log_2\left(1+ \frac{(1-a)\xi|h_2|^2}{a\xi|h_2|^2 + 1} \right)
\end{align}
When using regular NOMA, it is always true that if $|h_1|^2<|h_2|^2\rightarrow C_1(a)<C_{2\rightarrow 1}(a)$. However, since  CA-NOMA allows user-1 to subtract user-2's interference, this relationship of capacities no longer holds for CA-NOMA. Thus the effects of CA-NOMA on outage performances is investigated.

\section{CA-NOMA for Achieving a Minimum QoS Rate} \label{sec:QoS}

Suppose a downlink system requires a minimum rate $R_0$ in order to schedule any user's information to be transmitted at the base-station.  In a two-user CA-NOMA system, it is required that there are no outage states in the system, which includes the achievability of performing SIC. This gives rise to the following outage events
\begin{align}
    \label{eq:CA-NOMA_out1}C_1(a)<R_0& \Longrightarrow \mathcal{A}_1^\text{out}= \left\{|h_1|^2<\frac{2^{R_0}-1}{(1-a)\xi}\right\}\\
    \label{eq:CA-NOMA_out2}C_2(a)<R_0& \Longrightarrow \mathcal{A}_2^\text{out}= \left\{|h_2|^2<\frac{2^{R_0}-1}{a\xi} \right\}\\
    \label{eq:CA-NOMA_out21}C_{2\rightarrow 1}(a)<R_0& \Longrightarrow\mathcal{A}_{2\rightarrow 1}^\text{out}= \left\{|h_2|^2<\frac{2^{R_0}-1}{\xi(1-a2^{R_0})} \right\}.
\end{align}
If any of these outage events occurs, then the system will fail to perform a successful NOMA transmission. The union outage probability, which is a function of $a$, is then computed as 
\begin{align}
    p_\text{U-out}(a) &= \mathrm{Pr}\{ \mathcal{A}_1^\text{out}\cup\mathcal{A}_2^\text{out}\cup\mathcal{A}_{2\rightarrow 1}^\text{out} \} \nonumber\\
    & = 1 - \mathrm{Pr}\{ \overline{\mathcal{A}_1^\text{out}\cup\mathcal{A}_2^\text{out}\cup\mathcal{A}_{2\rightarrow 1}^\text{out}} \}\nonumber\\
    \label{eq:pout-union}&=1 - \mathrm{Pr}\{ \overline{\mathcal{A}_1^\text{out}}\cap\overline{\mathcal{A}_2^\text{out}}\cap\overline{\mathcal{A}_{2\rightarrow 1}^\text{out}} \}.
\end{align}

The most common and simplest approach to power allocation in NOMA systems is fixed-power allocation. However, instead of using an arbitrary fixed-power value, it is shown in the following subsection that there are fundamental restrictions on how fixed-power allocation should be used in CA-NOMA systems, and by extension all NOMA systems. 

\subsection{Fixed-power allocation  conditions for successful CA-NOMA} \label{subsec:fundamentals}
Suppose a fixed-power NOMA system is employed. The power allocation region required must be fundamentally explored. This power allocation region must be such that all of the events are feasible. It should be noted that there are channel condition cases such that it is not  possible for a user to achieve the minimum rate required by the system even with the optimum power allocation strategy, and thus no scheme exists that can guarantee meeting the system QoS requirement. 

By feasibility, it is meant that the power allocation region must be restricted such that the events $\mathcal{A}_1^\text{out}$, $\mathcal{A}_2^\text{out}$, and $\mathcal{A}_{2\rightarrow 1}^\text{out}$ occur with a probability less than 1. Therefore, the following property of NOMA, and by consequence CA-NOMA, is obtain.
\begin{property}\label{prop:certainoutage}
	The channel for user-2 to perform SIC is always in outage $\forall a>2^{-R_0}$ and $|h_2|^2>0$.
\end{property}
\begin{IEEEproof}
    The channel $h_2$ is in outage if
    \begin{align}
        &C_{2\rightarrow 1}(a) < R_0\\
        \Longrightarrow& \xi|h_2|^2(1-a2^{R_0})\leq 2^{R_0}-1.
    \end{align}
    However, since $\forall a>2^{-R_0}, \xi(1-a2^{R_0})<0$ 
    \begin{align}
		&\Longrightarrow |h_2|^2 < \frac{2^{R_0}-1}{\xi(1-a2^{R_0})} < 0.
	\end{align}
	which is impossible since $|h_2|^2>0$, therefore making the capacity impossible to be greater than $R_0$. Hence, 
	\begin{align}
        \forall |h_2|^2>0, \log_2\left(\frac{1+\xi|h_2|^2}{1+a\xi|h_2|^2} \right) < R_0.
    \end{align}
\end{IEEEproof}
\begin{remark}
	In general, if any user in a NOMA system must decode a signal in the presence of interference, where the interference cannot remove by SIC or caching, it will have this type of fundamental power allocation restriction due to the interference terms. Thus, in a NOMA system with minimum QoS rate requirement $R_0$, the power allocation coefficient of signals that will cause irremovable interference at certain receivers is fundamentally restricted by a function of the minimum rate $R_0$, independent of the channel gains. 
\end{remark}

According to property \ref{prop:certainoutage} of CA-NOMA, for any power allocation $a\leq 2^{-R_0}$,  there exists a non-zero probability that $C_{2\rightarrow 1}(a)>R_0$. Otherwise, SIC will fail, causing the entire purpose of using NOMA to fail. 

It is clear that there is a trade-off between $C_2(a)$ and $C_{2\rightarrow 1}(a)$ that depends on $a$, which can be evaluated using $\mathcal{A}_2^\text{out}$ and $\mathcal{A}_{2\rightarrow 1}^\text{out}$ as follows
\begin{align}
    &\frac{2^{R_0}-1}{a\xi} \lessgtr \frac{2^{R_0}-1}{\xi(1-a2^{R_0})}\\
    \label{eq:aub_fp}\Longrightarrow & \frac{1}{1+2^{R_0}}\lessgtr a.
\end{align}
This condition says that  if $a < \frac{1}{1+2^{R_0}}$, then $\overline{\mathcal{A}_2^\text{out}}\subset\overline{\mathcal{A}_{2\rightarrow 1}^\text{out}} \Longrightarrow \overline{\mathcal{A}_2^\text{out}}\cap\overline{\mathcal{A}_{2\rightarrow 1}^\text{out}} = \overline{\mathcal{A}_2^\text{out}}$. This implies that the probability in \eqref{eq:pout-union} can be simplified to 
\begin{align}
    \label{eq:puoutdefine}&p_\text{U-out}(a) = 1 - \mathrm{Pr}\{ \overline{\mathcal{A}_1^\text{out}}\cap\overline{\mathcal{A}_2^\text{out}} \}. 
\end{align}
Equation (\ref{eq:aub_fp}) also leads to the following.
\begin{lemma}\label{lemma:fp_outage}
	Let $\mathcal{A}_{2, \text{oma}}^\text{out} = \left\{|h_2|^2: \frac{1}{2}\log_2(1+\xi|h_2|^2) <R_0\right\}$. Then $\forall a\leq\frac{1}{1+2^{R_0}}$,
	\begin{align}
		\label{eqsubsetout2}\mathcal{A}_{2\rightarrow 1}^\text{out}\subseteq\mathcal{A}_{2, \text{oma}}^\text{out}\subseteq\mathcal{A}_2^\text{out}.
	\end{align}
\end{lemma}
\begin{IEEEproof} 
	Define the functions $B_2(a) = \frac{2^{R_0}-1}{a\xi}$ and $B_{2\rightarrow 1}(a)=\frac{2^{R_0}-1}{\xi(1-a2^{R_0})}$. $B_2(a)$ is a montonically decreasing function of $a\in(0,1)$, while $B_{2\rightarrow 1}(a)$ is a monotonically increasing function of $a\in(0,2^{-R_0})$. If $a\leq\frac{1}{1+2^{R_0}}$, then
	\begin{align*}
		\label{eq:monotonicC2}B_2(a)\geq B_2(\tfrac{1}{1+2^{R_0}})=B_{2\rightarrow 1}(\tfrac{1}{1+2^{R_0}})\geq B_{2\rightarrow 1}(a)
	\end{align*}
	Let $\mathcal{A}_2^{\text{out}*} = \{|h_2|^2: |h_2|^2< B_2(\frac{1}{1+2^{R_0}})  \}$, so $B_2(\frac{1}{1+2^{R_0}})\leq B_2(a)\Rightarrow \mathcal{A}_2^{\text{out}*} \subseteq \mathcal{A}_{2}^\text{out},\forall a\leq\frac{1}{1+2^{R_0}}$. However, the event 
	\begin{align*}
		\mathcal{A}_2^{\text{out}*} =&\{|h_2|^2: |h_2|^2<B_2(\tfrac{1}{1+2^{R_0}}) \}\\
			=&\{|h_2|^2: |h_2|^2 < \tfrac{(2^{R_0}-1)(1+2^{R_0})}{\xi} \}\\
			=&\{|h_2|^2:|h_2|^2< \tfrac{4^{R_0}-1}{\xi}\}\\
			=&\{|h_2|^2:\tfrac{1}{2}\log_2(1+\xi|h_2|^2)<R_0\} = \mathcal{A}_{2, \text{oma}}^\text{out}
	\end{align*}
	Therefore, since $\mathcal{A}_2^{\text{out}*}=\mathcal{A}_{2, \text{oma}}^\text{out}\Rightarrow \mathcal{A}_{2, \text{oma}}^\text{out} \subseteq \mathcal{A}_2^{\text{out}}$.

	Similarly, let $\mathcal{A}_{2\rightarrow 1}^{\text{out}*} = \{|h_2|^2: |h_2|^2< B_{2\rightarrow 1}(\frac{1}{1+2^{R_0}}) \}$, so $B_{2\rightarrow 1}(\frac{1}{1+2^{R_0}})\geq B_{2\rightarrow 1}(a)\Rightarrow \mathcal{A}_{2\rightarrow 1}^\text{out}\subseteq \mathcal{A}_{2\rightarrow 1}^{\text{out}*} ,\forall a\leq\frac{1}{1+2^{R_0}}$. Note that the event 
	\begin{align*}
		&\mathcal{A}_{2\rightarrow 1}^{\text{out}*} = \{ |h_2|^2: |h_2|^2< B_{2\rightarrow 1}(\tfrac{1}{1+2^{R_0}}) \}\\
     &=\left\{|h_2|^2: |h_2|^2<\frac{2^{R_0}-1}{\xi(1-\frac{2^{R_0}}{1+2^{R_0}})} \right\}\\
			&=\{|h_2|^2: |h_2|^2<\tfrac{(2^{R_0}-1)(1+2^{R_0})}{\xi} \}\\
     &=\{|h_2|^2: |h_2|^2<\tfrac{4^{R_0}+1}{\xi}\}\\
			&=\{|h_2|^2: \tfrac{1}{2}\log_2(1+\xi|h_2|^2)<R_0\} = \mathcal{A}_{2, \text{oma}}^\text{out}
	\end{align*}
	Therefore, since $\mathcal{A}_{2\rightarrow 1}^{\text{out}*}=\mathcal{A}_{2, \text{oma}}^\text{out}\Rightarrow \mathcal{A}_{2\rightarrow 1}^{\text{out}} \subseteq \mathcal{A}_{2, \text{oma}}^\text{out}$. Hence, $\mathcal{A}_{2\rightarrow 1}^\text{out}\subseteq\mathcal{A}_{2, \text{oma}}^\text{out}\subseteq\mathcal{A}_2^\text{out},\forall a\leq\frac{1}{1+2^{R_0}}$.
\end{IEEEproof}

With lemma \ref{lemma:fp_outage}, it is clear that fundamentally the power allocation region of $a$ is restricted to the set $(0,\frac{1}{1+2^{R_0}}]$ in order to minimize the union-outage probability according to equation (\ref{eq:puoutdefine}), which is done in the following subsection.

\subsection{Outage Probabilities} \label{subsec:poutage}
The joint p.d.f. of $|h_1|^2$ and $|h_2|^2$ for users paired randomly is given\footnote{Since the possession of a cache is independent of the order of the channel gains, the distribution of the channel gains is the same as  in \cite{FairNOMAFull}} \cite{FairNOMAFull} by $f_{|h_1|^2,|h_2|^2}(x_1,x_2) = \frac{2}{\beta^2}e^{-\frac{x_1+x_2}{\beta}}$, for $x_1,x_2\geq 0$. Since $\mathrm{Pr}\{ \overline{\mathcal{A}_1^\text{out}}\cap\overline{\mathcal{A}_2^\text{out}} \} = \mathrm{Pr}\{ \overline{\mathcal{A}_1^\text{out}} \} - \mathrm{Pr}\{ \overline{\mathcal{A}_1^\text{out}}\cap\mathcal{A}_2^\text{out} \}$, and $a<\frac{1}{2}\Rightarrow\frac{2^{R_0}-1}{(1-a)\xi} < \frac{2^{R_0}-1}{a\xi}$, then computing
\begin{align}
    \mathrm{Pr}\{ \overline{\mathcal{A}_1^\text{out}} \} &= \int_{\frac{2^{R_0}-1}{(1-a)\xi}}^\infty \int_{x_1}^\infty \frac{2}{\beta^2}e^{-\frac{x_1+x_2}{\beta}}dx_2 dx_1 = e^{-\frac{2(2^{R_0}-1)}{(1-a)\beta\xi}},
\end{align}
and 
\begin{align}
     &\mathrm{Pr}\{ \overline{\mathcal{A}_1^\text{out}}\cap\mathcal{A}_2^\text{out} \} = \int_{\frac{2^{R_0}-1}{(1-a)\xi}}^{\frac{2^{R_0}-1}{a\xi}} \int_{\frac{2^{R_0}-1}{(1-a)\xi}}^{x_2}\frac{2}{\beta^2}e^{-\frac{x_1+x_2}{\beta}}dx_1 dx_2 \nonumber\\
     &	= -2e^{-\frac{2^{R_0}-1}{a(1-a)\beta\xi}} + 2e^{-\frac{2(2^{R_0}-1)}{(1-a)\beta\xi}} +e^{-\frac{2(2^{R_0}-1)}{a\beta\xi}} - e^{-\frac{2(2^{R_0}-1)}{(1-a)\beta\xi}},
\end{align}
gives the probability of union-outage as
\begin{align}
    &p_\text{U-out}(a) = 1 + \mathrm{Pr}\{ \overline{\mathcal{A}_1^\text{out}}\cap\mathcal{A}_2^\text{out} \} - \mathrm{Pr}\{ \overline{\mathcal{A}_1^\text{out}} \}\nonumber\\
	\label{eq:p_uout_closedform}&= 1 + e^{-\frac{2(2^{R_0}-1)}{a\beta\xi}} - 2e^{-\frac{2^{R_0}-1}{a(1-a)\beta\xi}}.
\end{align}
The following lemma provides a tight approximation for the optimum power allocation coefficient that minimizes equation (\ref{eq:p_uout_closedform}).
\begin{lemma}\label{lem:amin}
	The power allocation coefficient $a_\text{min}\in(0,\frac{1}{1+2^{R_0}}]$ that minimizes $p_\text{U-out}(a)$ is given approximately by
    \begin{align}
        a_\text{min} \approx \min\left\{\tfrac{A_0}{3} + \left(\tfrac{A_1+A_2}{2}\right)^{\frac{1}{3}} -\left(\tfrac{-A_1+A_2}{2}\right)^{\frac{1}{3}}, \tfrac{1}{1+2^{R_0}}\right\},
    \end{align}
    where $A_0 = \frac{2^{R_0}-1}{\beta\xi}$, $A_1 = \frac{2}{27}A_0^3-\frac{2}{3}A_0^2+A_0$, and $A_2 = A_0\sqrt{1-\frac{4}{27}A_0}$. 
\end{lemma}
\begin{IEEEproof}
    Using the power series for $t\ll 1$
    \begin{equation}
        e^t = \sum_{k=0}^\infty \frac{t^k}{k!} = 1+t +O(t^2),
    \end{equation}
    $p_\text{U-out}(a)$ can be approximated as
    \begin{align}
        p_\text{U-out}(a) \approx& 1 + \left(1 - \frac{2^{R_0}-1}{a\beta\xi} \right)^2 - 2\left(1 - \frac{2^{R_0}-1}{a(1-a)\beta\xi} \right)\nonumber\\
        \label{eq:pout_approx}=&\frac{A_0^2}{a^2}+\frac{2A_0}{1-a}.
    \end{align}
    Taking the derivative of (\ref{eq:pout_approx}) with respect to $a$, equating it to $0$, and solving for $a$ gives the following
    \begin{align}
        \label{eq:dpoutda}&\frac{d}{da}p_\text{U-out}(a) = -\frac{2A_0^2}{a^3}+\frac{2A_0}{(1-a)^2}=0\\
        &\Longrightarrow 0 = a^3-A_0a^2+2A_0a-A_0.
    \end{align}
    The solution to the cubic is as follows \cite{cubic_sol}. Let $a = b+\frac{A_0}{3}$. Then the cubic equation becomes
    \begin{align}
    	0=&\left(b+\frac{A_0}{3}\right)^3-A_0\left(b+\frac{A_0}{3}\right)^2+2A_0\left(b+\frac{A_0}{3}\right)-A_0\\
    	\label{eq:cubic1}\Longrightarrow & b^3 + b\left(2A_0-\tfrac{1}{3}A_0^2\right) = \tfrac{2}{27}A_0^3-\tfrac{2}{3}A_0^2+A_0.
    \end{align}
    Let $A_1 = \frac{2}{27}A_0^3-\frac{2}{3}A_0^2+A_0$ and $A_3 = 2A_0-\frac{1}{3}A_0^2$. Then equation (\ref{eq:cubic1}) looks like $b^3+A_3b = A_1$. Set the numbers $q$ and $s$ to be defined by $3sq = A_3$ and $s^3-q^3=A_1$. Note that $b=s-q$ solves equation (\ref{eq:cubic1}). Therefore, since $s=\frac{A_3}{3q}$,
    \begin{align}
    	&\left(\frac{A_3}{3q}\right)^3 - q^3=A_1\\
    	\Longrightarrow& 0=q^6+A_1q^3-\tfrac{1}{27}A_3^3\\
    	\Longrightarrow& q = \left(\frac{-A_1+\sqrt{ A_1^2 + \tfrac{4}{27}A_3^3}}{2}\right)^{\frac{1}{3}}\\
    				&=\left(\frac{-A_1+A_2}{2}\right)^{\frac{1}{3}},
    \end{align}
	where $\sqrt{A_1^2 + \tfrac{4}{27}A_3^3} = A_0\sqrt{1-\frac{4}{27}A_0}= A_2$. Then $s$ is found by
    \begin{equation}
    	s = (q^3+A_1)^\frac{1}{3} = \left(\frac{A_1+A_2}{2}\right)^{\frac{1}{3}}.
    \end{equation}
    The value of $a$ that makes equation (\ref{eq:dpoutda}) true is 
    \begin{align}
    	a &=\frac{A_0}{3}+ b =\frac{A_0}{3} + s-t\\
    		&=\frac{A_0}{3}+\left(\frac{A_1+A_2}{2}\right)^{\frac{1}{3}}-\left(\frac{-A_1+A_2}{2}\right)^{\frac{1}{3}}.
    \end{align}
    Therefore, since $a\leq (1+2^{R_0})^{-1}$, then 
    \begin{equation}
    	a_\text{min} \approx \min\left\{\tfrac{A_0}{3} + \left(\tfrac{A_1+A_2}{2}\right)^{\frac{1}{3}} -\left(\tfrac{-A_1+A_2}{2}\right)^{\frac{1}{3}}, \tfrac{1}{1+2^{R_0}}\right\}.
    \end{equation}
\end{IEEEproof}
Lemma \ref{lem:amin} provides  a  close approximation to the optimum power allocation coefficient that minimizes the outage probability when using fixed-power allocation strategies in CA-NOMA systems. Furthermore, $a_\text{min}$ becomes strictly less than $\frac{1}{1+2^{R_0}}$ as $\xi$ increases. This is beneficial because user-2's strong channel and ability to achieve $C_2(a)>R_0$ allows the system to then allocate more power to the weaker user to also achieve the minimum rate,  while making the success of performing SIC guaranteed if $C_2(a)>R_0$.

\section{Comparison of theoretical and simulation results} \label{sec:results}
\begin{figure}
	\centering
	\includegraphics[scale=0.6]{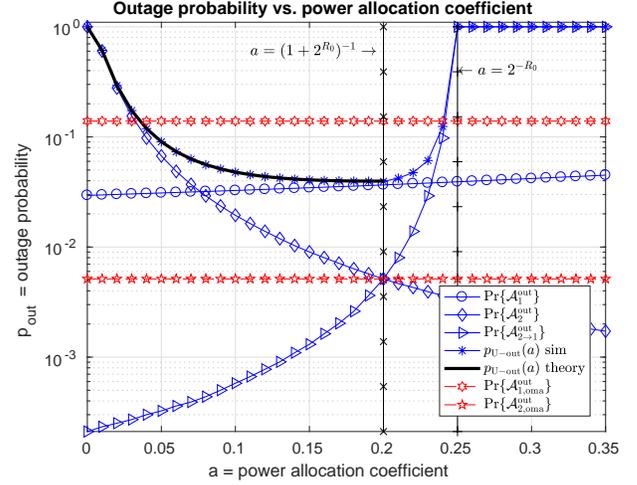}
	\caption{\label{fig:puout_vs_a}Outage probabilities are plotted vs. $a$; $\xi=20$ dB, $\beta=2$, $R_0=2$ bps/Hz}
\end{figure}

The plot in figure \ref{fig:puout_vs_a} demonstrates the different outage probabilities for each user. Note that the result for $p_\text{U-out}(a)$ in equation (\ref{eq:p_uout_closedform}) exactly matches the simulation for $a\in(0,\frac{1}{1+2^{R_0}}]$, which is the interval of interest, given Lemma \ref{lemma:fp_outage}. When $a\in(\frac{1}{1+2^{R_0}}, 2^{-R_0})$, $p_\text{U-out}(a)$ is dominated by $\mathrm{Pr}\{ \mathcal{A}_{2\rightarrow 1}^\text{out} \}$, which is undesirable since SIC must happen in order to have a functioning NOMA system. When $a\geq 2^{-R_0}$, $p_\text{U-out}(a)=\mathrm{Pr}\{ \mathcal{A}_{2\rightarrow 1}^\text{out} \}=1$, which simply means that SIC is impossible, causing the NOMA system to surely fail. 

\begin{figure}
	\centering
	\includegraphics[scale=0.6]{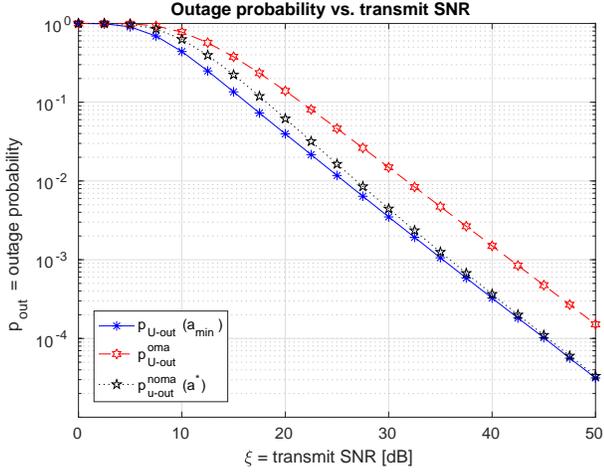}
	\caption{\label{fig:puout_vs_snr}Outage probabilities are plotted vs. $\xi$, with $a=a_\text{min}$; $\beta=2$, $R_0=2$ bps/Hz}
\end{figure}

In figure \ref{fig:puout_vs_snr}, the plot of $p_\text{U-out}(a)$ vs. $\xi$ is demonstrated for the minimizing $a_\text{min}$, and compared with the union-outage probabilities of OMA and regular NOMA, where regular NOMA has power allocation coefficient $a^*$ that minimizes its union-outage probability. It can be seen that the union-outage probability using CA-NOMA is superior to both regular NOMA and OMA. The difference is most noticible when $10<\xi<30$ dB. However, as $\xi$ increases, regular NOMA performs almost as well. This makes sense because a very large $\xi$ will make outage very unlikely, given that regular NOMA still allows the flexibility of finding an optimum $a^*$. 

\begin{figure}
	\centering
	\includegraphics[scale=0.6]{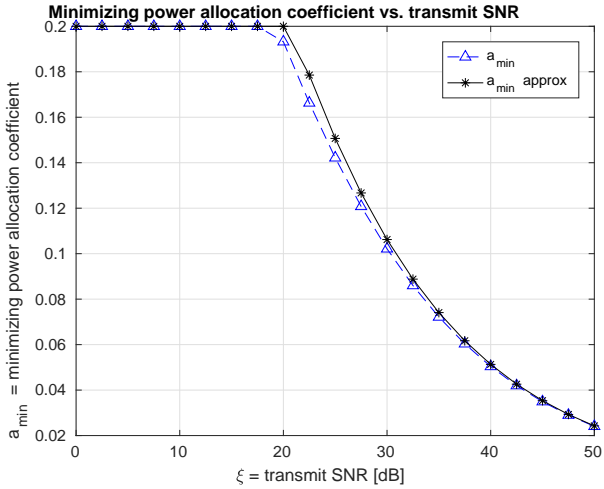}
	\caption{\label{fig:amin}$a_\text{min}$ approximation compared to actual simulated value; $R_0=2$ bps/Hz, $\beta=2$}
\end{figure}

Figure \ref{fig:amin} demonstrates the tightness of the approximation of $a_\text{min}$ given by lemma \ref{lem:amin}. It is worth noting that there is a threshold where $a_\text{min}<\frac{1}{1+2^{R_0}}$ that is a function of $\xi$ and $2^{R_0}$. As expected, the value of $a_\text{min}$ decreases as $\xi$ increases, which allows the system to take advantage of the strong channel to help the weak channel have greater probability of achieving the minimum QoS rate. From the results in figures \ref{fig:puout_vs_snr} and \ref{fig:amin}, it is clear that the union-outage performance using CA-NOMA is better than the performance of OMA and regular NOMA. 



\section{Conclusion and Future Work}\label{sec:conclusion}
Using the two candidate technologies of NOMA and caching at the user terminals is demonstrated to help a system improve probability of achieving the mimimum QoS rate required by a system. Using these results, it is clear that any NOMA system can take advantage of users that  possess caches of content of other users in the system. The resulting CA-NOMA system can further be investigated to determine how the power allocation coefficients are affected when the number of simultaneous NOMA users is more than 2. Furthermore, performance of CA-NOMA can be improved upon by fundamentally determining the power allocation coefficients of users by utilizing their channel gains in their derivation similar to \cite{FairNOMAFull}. 

The benefits of CA-NOMA can be extended to MIMO systems, where a users in different clusters (and thus allocated along different degrees-of-freedom) can have their interference eliminated at receivers if a content is cached. Furthermore, the distribution of cached content based on its popularity can be used to determine how often CA-NOMA can be fully utilized on top of regular NOMA in multi-user systems.

\end{document}